\documentclass[aps,prb,preprint,showpacs]{revtex4}
\usepackage{graphicx}

\begin{document}
\sloppy
\title{Kink structure in the electronic dispersion of high-$T_{\mathrm c}$ 
superconductors from the electron-phonon interaction}
\author{Shigeru Koikegami}
\email[email address: ]{shigeami@h3.dion.ne.jp }
\affiliation{Second Lab, LLC, 10-7-204 Inarimae, Tsukuba 305-0061, Japan \\
Nanoelectronics Research Institute, 
AIST Tsukuba Central 2, Tsukuba 305-8568, Japan}
\author{Yoshihiro Aiura}
\affiliation{Nanoelectronics Research Institute, 
AIST Tsukuba Central 2, Tsukuba 305-8568, Japan}
\date{today}

\begin{abstract}
We investigate the electronic dispersion of high-$T_{\mathrm c}$ 
superconductor on the basis of the two-dimensional three-band Hubbard model 
with the electron-phonon interaction 
together with the strong electron-electron interaction.
In our model, it is shown across the hole-doped region of high-$T_{\mathrm c}$ 
superconductor that the electron-phonon interaction makes a dispersion kink,
observed along the nodal direction, and that the small isotope effect
appears on the electronic dispersion. 
\end{abstract}


\maketitle

\section*{INTRODUCTION}
For the past two decades, extensive studies of 
high-$T_{\mathrm c}$ cuprates have spotlighted many curious phenomena. 
It has been argued that most phenomena are 
attributable to the strong correlations 
among electrons, which play 
significant roles in these materials. However, 
since the discovery of sudden changes in the electron dispersion 
or ``kinks'' shown by the 
angle-resolved photoemission spectroscopy 
(ARPES),~\cite{Kaminski01,Lanzara01} 
effects of electron-boson interactions on electronic self-energies 
have been recognized. 
While the kinks are now indisputable in cuprates,~\cite{Zhou03} 
their origin as arising from electronic coupling to phonons~\cite{ZXShen02} 
or magnetic excitations~\cite{Terashima06,Zabolotnyy06,Kordyuk06} 
remains unclear. 

Recent scanning tunneling microscope study 
showed that the statistical distribution of energy of bosonic modes   
($\Omega$) has meaningful difference between 
$^{16}{\mathrm O}$ and $^{18}{\mathrm O}$ materials. 
Thus, it should be hard to exclude the 
possibility that electron-phonon interaction (EPI) 
significantly affects the electronic states in cuprates. 

In this study, we investigate the analysis upon the EPI together with 
the electron-electron interaction (EEI) on the basis of 
the two-dimensional (2D) three-band Hubbard--Holstein (HH) model. 
With the use of our three-band HH model, we can reproduce the situation of 
the real high-$T_{\mathrm c}$ materials, 
in which EPI mainly works on $p$ electrons at O sites. 

\section*{FORMULATION}

Our model Hamiltonian $H$ is composed of 
$d$ electrons at each Cu site, $p$ electrons at O site, 
and lattice vibrations of O atoms. 
We consider only the on-site Coulomb repulsion $U$ between 
$d$ electrons at each Cu site as our EEI. 
Let us define that $\mu$ and $N_0$ represent the chemical potential and 
the number of all electrons, respectively. 
Then, $H-\mu N_0$ is divided into the non-interacting part, $H_0$, 
the electron-electron interacting part $H_{\mathrm{el-el}}$, 
the phonon part $H_{\mathrm{ph}}$, 
and the electron-phonon interacting part $H_{\mathrm{el-ph}}$ as
\begin{eqnarray}
H - \mu N_0 & = & H_0+H_{el-el}+H_{ph}+H_{el-ph}, \nonumber \\
        N_0 & = & \sum_{{\mathbf k} \sigma}
(d_{{\mathbf k} \sigma}^{\dagger}d_{{\mathbf k} \sigma}
+p_{{\mathbf k} \sigma}^{x \dagger}p_{{\mathbf k} \sigma}^x
+p_{{\mathbf k} \sigma}^{y \dagger}p_{{\mathbf k} \sigma}^y).
\end{eqnarray} 
Here, $d_{{\mathbf k} \sigma}(d_{{\mathbf k} \sigma}^{\dagger})$ and 
$p_{{\mathbf k} \sigma}^{x(y)}(p_{{\mathbf k} \sigma}^{x(y) \dagger})$ are 
the annihilation (creation) operator for $d$ and 
$p^{x(y)}$ electrons of momentum ${\mathbf k}$ and spin $\sigma$, 
respectively. The non-interacting part $H_0$ is represented by 
\begin{eqnarray}
H_0 & = & \sum_{{\mathbf k} \sigma}
\left(
d_{{\mathbf k} \sigma}^\dagger \, 
p_{{\mathbf k} \sigma}^{x \dagger} \, 
p_{{\mathbf k} \sigma}^{y \dagger}
\right) \left(
\begin{array}{ccc} \Delta_{dp} & \zeta_{\mathbf k}^x  & \zeta_{\mathbf k}^y \\ 
-\zeta_{\mathbf k}^x & 0 & \zeta_{\mathbf k}^p \\
-\zeta_{\mathbf k}^y & \zeta_{\mathbf k}^p & 0 \\
\end{array} \right)\!
\left( \begin{array}{c} d_{{\mathbf k} \sigma} \\ p_{{\mathbf k} \sigma}^x
\\ p_{{\mathbf k} \sigma}^y \\
\end{array} \right) \nonumber \\
    & \equiv & {\sum_{{\mathbf k} \sigma}}
\mathbf{d}_{{\mathbf k} \sigma}^\dagger \, 
\mathbf{H}_0\,
\mathbf{d}_{{\mathbf k} \sigma}.
\label{eq:6}
\end{eqnarray}
We take the lattice constant of the square
lattice formed from Cu sites as the unit of length. Then, $\zeta_{\mathbf k}^{x(y)}=2{\rm i}\,t_{dp} \sin \frac{k_{x(y)}}{2}$ and
$\zeta_{\mathbf k}^p=-4t_{pp} \sin \frac{k_x}{2} \sin \frac{k_y}{2}$,
where $t_{dp}$ is the transfer energy between a $d$ orbital and a
neighboring $p^{x(y)}$ orbital and $t_{pp}$ is that between a
$p^x$ orbital and a $p^y$ orbital. 
$\Delta_{dp}$ is the difference of energy levels of $d$ and $p$ orbitals. 
In this study, we take $t_{dp}$ as the unit of energy. The residual parts are 
described as follows:
\begin{equation}
H_{el-el}= \frac{U}{N} \sum_{{\mathbf k} {\mathbf k}^\prime} 
        \sum_{{\mathbf q}(\neq{\mathbf 0})}
        d_{{\mathbf k}+{\mathbf q} \uparrow}^{\dagger} d_{{\mathbf
        k}^\prime- {\mathbf q} \downarrow}^{\dagger} d_{{\mathbf k}^\prime
        \downarrow} d_{{\mathbf k} \uparrow},
\end{equation}
\begin{equation}
H_{ph}= \sum_{{\mathbf q}}\!\sum_{\nu=\{x,y\}} \omega^\nu_{{\mathbf q}}\,
        b^{\nu \dagger}_{{\mathbf q}} b^\nu_{{\mathbf q}},
\end{equation}
and
\begin{equation}
H_{el-ph}= \frac{1}{N} \sum_{{\mathbf k} \sigma} 
        \hspace{4pt}\sum_{{\mathbf q}}\!\sum_{\nu=\{x,y\}}
	g\,\alpha^\nu_{\mathbf{k},\mathbf{q}}\,
        p^{\nu \dagger}_{{\mathbf k}+{\mathbf q}\sigma}
        p^\nu_{{\mathbf k} \sigma}
 \left( b^\nu_{{\mathbf q}}+b^{\nu \dagger}_{-{\mathbf q}} \right),
\label{eq:8}
\end{equation}
where $U$ is the on-site Coulomb repulsion between $d$ orbitals, 
$N$ is the number of ${\mathbf k}$-space lattice points in the first
Brillouin zone (FBZ), and $g\,\alpha^\nu_{\mathbf{k},\mathbf{q}}
\,(\nu=\{x,y\})$ 
is the electron-phonon matrix element, respectively. 
We consider that the half-breathing phonon mode,~\cite{McQueeney01} 
in which oxygen ions are vibrating 
along the $x$ or $y$ directions, is crucial for our problem. 
Thus, ignoring the other phonon modes, 
we have the electron-phonon interacting part as Eq.~(\ref{eq:8}).

Then, we introduced the unperturbed and 
perturbed Green's functions, which are to be described in $3\times3$ 
matrix form. The unperturbed Green's function 
$\mathbf{G}^{(0)}(\mathbf{k},z)$ is derived from Eq.~(\ref{eq:6}) as
\begin{equation}
\mathbf{G}^{(0)}(\mathbf{k},z)=\left[z\mathbf{I}-\mathbf{H}_0\right]^{-1},
\end{equation}
where $\mathbf{I}$ is a $3\times3$ unit matrix. 
Using the abbreviation of Fermion Matsubara frequencies, 
$\epsilon_n=\pi T(2n+1)$ with integer $n$ and temperature $T$, 
the perturbed Green function $\mathbf{G}(\mathbf{k},z)$ 
is determined by the Dyson equation:
\begin{equation}
\mathbf{G}(\mathbf{k},\mathrm{i}\epsilon_n)\,^{-1}=
\mathbf{G}^{(0)}(\mathbf{k},\mathrm{i}\epsilon_n)\,^{-1}
-\mathbf{\Sigma}(\mathbf{k},\mathrm{i}\epsilon_n),
\label{eq:1}
\end{equation}
where $\mathbf{\Sigma}(\mathbf{k},\mathrm{i}\epsilon_n)$ is the self-energy 
expected to be a diagonal matrix. 
In order to estimate the $d$ electron self-energy  
$\Sigma_d(\mathbf{k},\mathrm{i}\epsilon_n) \equiv 
\Sigma_{11}(\mathbf{k},\mathrm{i}\epsilon_n)$ in Eq.~(\ref{eq:1}), 
we adopt the fluctuation exchange approximation~\cite{Bickers89} 
as follows:
\begin{equation}
\Sigma_d(\mathbf{k},\mathrm{i}\epsilon_n) = \frac{T}{N}
\sum_{\mathbf{q}\,m}
G_d(\mathbf{k}-\mathbf{q},\mathrm{i}\epsilon_n-\mathrm{i}\omega_m)\,
V_{\mathrm{el-el}}(\mathbf{q},\mathrm{i}\omega_m),
\label{eq:2}
\end{equation}
where $G_d(\mathbf{k},\mathrm{i}\epsilon_n) \equiv G_{11}(\mathbf{k},\mathrm{i}\epsilon_n)$,
\begin{eqnarray}
\label{eq:5}
V_{\mathrm{el-el}}(\mathbf{q},\mathrm{i}\omega_m) 
& = & \frac{3}{2}\frac{U^2\chi(\mathbf{q},\mathrm{i}\omega_m)}
{1-U\chi(\mathbf{q},\mathrm{i}\omega_m)}+\frac{1}{2}\frac{U^2\chi(\mathbf{q},\mathrm{i}\omega_m)}
{1+U\chi(\mathbf{q},\mathrm{i}\omega_m)} \nonumber \\
&   & -U^2\chi(\mathbf{q},\mathrm{i}\omega_m),
\end{eqnarray}
where $\omega_m=2m\,\pi T$ with integer $m$ are Boson Matsubara frequencies, 
and
\begin{equation}
\label{eq:7}
\chi(\mathbf{q},\mathrm{i}\omega_m)= -\frac{T}{N}\sum_{\mathbf{k}\,n}
G_d(\mathbf{q}+\mathbf{k},\mathrm{i}\omega_m+\mathrm{i}\epsilon_n)
\,G_d(\mathbf{k},\mathrm{i}\epsilon_n).
\end{equation}
In order to estimate the p$^{x(y)}$-electron self-energy 
$\Sigma_{x(y)}(\mathbf{k},\mathrm{i}\epsilon_n) \equiv 
\Sigma_{22(33)}(\mathbf{k},\mathrm{i}\epsilon_n)$ in Eq.~(\ref{eq:1}), 
we exploit the Brillouin-Wigner perturbation theory. 
We adopt the self-consistent one-loop approximation as follows:
\begin{equation}
\Sigma_{x(y)}(\mathbf{k},\mathrm{i}\epsilon_n) = \frac{T}{N}
\sum_{\mathbf{q}\,m}
G_{x(y)}(\mathbf{k}-\mathbf{q},\mathrm{i}\epsilon_n-\mathrm{i}\omega_m)\,
V_{\mathrm{el-ph}}^{x(y)}(\mathbf{q},\mathrm{i}\omega_m),
\label{eq:3}
\end{equation}
where $G_{x(y)}(\mathbf{k},\mathrm{i}\epsilon_n) 
\equiv G_{22(33)}(\mathbf{k},\mathrm{i}\epsilon_n)$ and 
$V_{\mathrm{el-ph}}^{x(y)}(\mathbf{q},\mathrm{i}\omega_m)$ is the EPI on p$^{x(y)}$ electron. 
Our EPI is determined as follows:
\begin{equation}
V_{\mathrm{el-ph}}^{x(y)}(\mathbf{q},\mathrm{i}\omega_m) = 
\lambda \left|\alpha^{x(y)}_{\mathbf{q}}\right|^{\,2}
\left[\frac{1}{\omega_{\mathrm{h}}+\mathrm{i}\omega_m}+\frac{1}{\omega_{\mathrm{h}}-\mathrm{i}\omega_m}\right],
\label{eq:9}
\end{equation}
where $\lambda=g^2/(2\omega_{\mathrm{h}})$ 
and $\alpha^{x(y)}_{\mathbf{q}}=\sin \frac{q_{x(y)}}{2}$. 
$\omega_{\mathrm{h}}$ is 
the specific phonon energy for the half-breathing mode. As above, 
we ignore the effects in which EEI and EPI are coupled. Thus, as shown 
by Eqs.~(\ref{eq:2})-(\ref{eq:9}), in our formulation, EEI and EPI are completely decoupled. 
Of course, this assumption is inadequate to 
analyze the case in which the characteristic 
energy due to EEI is comparable to the phonon energy. 
However, as will be seen later, we actually treat the cases with rather high 
characteristic energy due to EEI. Hence, decoupling EEI with EPI 
should be justified in our analysis.
The electron-phonon coupling constant $\lambda$ does not depend on 
$M_{\mathrm{O}}$ since $\omega_{\mathrm{h}} \propto M_{\mathrm{O}}^{-1/2}$ and 
$g \propto \left(M_{\mathrm{O}}\omega_{\mathrm{h}}\right)^{-1/2}$, 
where $M_{\mathrm{O}}$ is the mass of an oxygen ion. 
In our model, thus, the isotope effect is reflected on the phonon energy in 
the electronic self-energies only, but not any changes in the strength. 

\section*{RESULTS AND DISCUSSION}

We need to solve 
Eqs.~(\ref{eq:1})--(\ref{eq:3}) 
in a fully self-consistent manner. 
During numerical calculations, 
we divide the FBZ into $128 \times 128$ meshes. We prepare $2^{12}=4096$ 
Matsubara frequencies for temperature $T \sim 87{\mathrm{K}}$. 
As shown later, at this temperature, our calculation can reproduce 
the important behavior of electrons in normal state. Moreover, to our knowledge, 
the situation will not be changed if we change $T$ to some extent. 

$t_{dp} \sim 1.0\,{\mathrm{eV}}$ and 
$t_{pp} \sim 0.55\,{\mathrm{eV}}$, which are all common for our calculations. 
These values are chosen so that we can reproduce 
the typical Fermi surface of Bi$_2$Sr$_2$CaCu$_2$O$_{8+\delta}$ 
observed by ARPES.~\cite{Feng2001,Chuang2001} 
$\Delta_{dp} \sim 1.4\,{\mathrm{eV}}$, 
$U \sim 3.0\,{\mathrm{eV}}$, and $\lambda=0.8$ unless stated.
The phonon energy is set as $\omega_{\mathrm{h}} \sim 65(61)\,{\mathrm{meV}}$ for 
$^{16}{\mathrm O}(^{18}{\mathrm O})$ material.
\begin{table}
\begin{ruledtabular}
\begin{tabular}{ccccc}
Isotope & $n_d^h$ & $n_p^h$ & $\delta$ & \\ \hline
  $^{16}{\mathrm O}$ & $0.5575$ & $0.4837$ & $0.0412$ & LD \\
  $^{18}{\mathrm O}$ & $0.5574$ & $0.4838$ & $0.0412$ & LD \\
  $^{16}{\mathrm O}$ & $0.5932$ & $0.5039$ & $0.0971$ & UD \\
  $^{18}{\mathrm O}$ & $0.5931$ & $0.5040$ & $0.0972$ & UD \\
  $^{16}{\mathrm O}$ & $0.6281$ & $0.5258$ & $0.1540$ & OP \\
  $^{18}{\mathrm O}$ & $0.6281$ & $0.5259$ & $0.1540$ & OP \\
  $^{16}{\mathrm O}$ & $0.6683$ & $0.5525$ & $0.2209$ & OD \\
  $^{18}{\mathrm O}$ & $0.6683$ & $0.5526$ & $0.2209$ & OD \\
\end{tabular}
\end{ruledtabular}
\caption{\label{table:1}Number of doped holes. $\delta \equiv n_d^h+n_p^h-1$.}
\end{table}
\begin{figure}
\includegraphics[width=5.2cm]{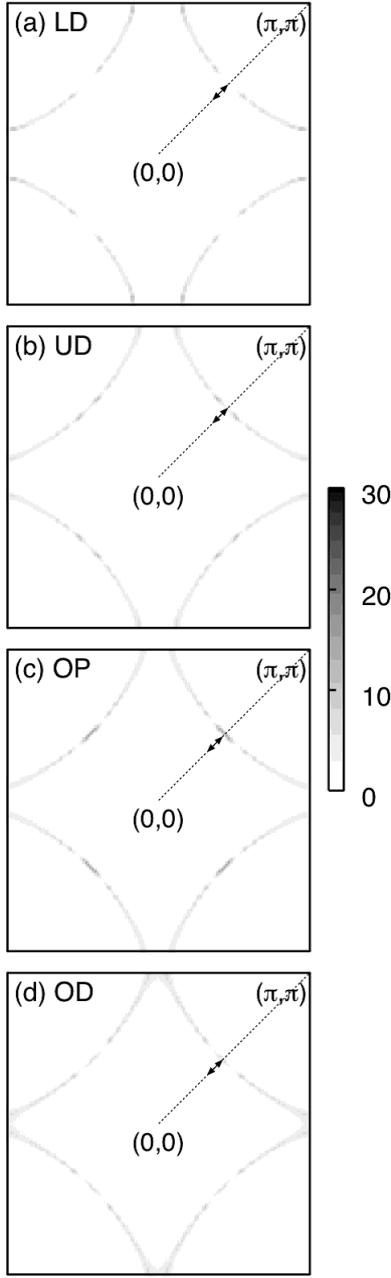}
\caption{\label{figure:1}Fermi surfaces for $^{16}{\mathrm O}$ materials 
indicated in Table~\ref{table:1}.}
\end{figure}
\begin{figure}
\includegraphics[width=8.4cm]{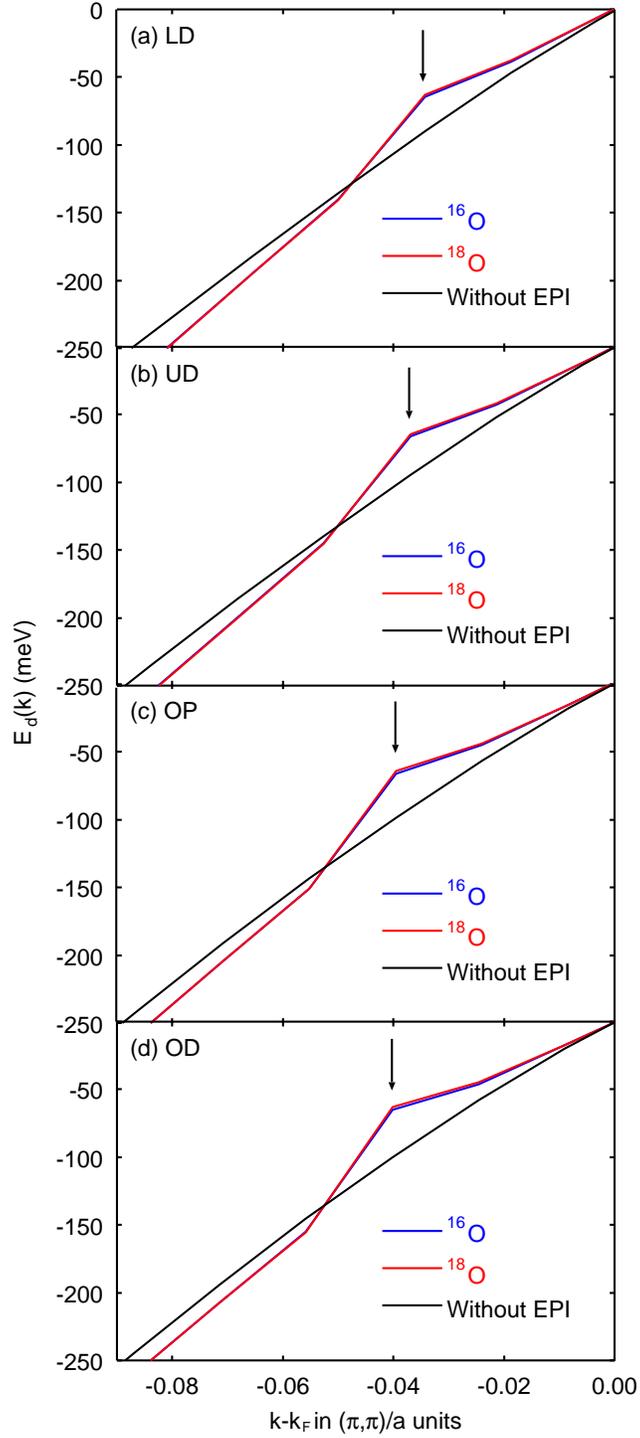}
\caption{\label{figure:2}Dispersion kinks along the nodal direction. 
Momentum is measured from each Fermi surface. Arrows indicate the momenta 
at which kinks occur.}
\end{figure}
\begin{figure}
\includegraphics[width=8.4cm]{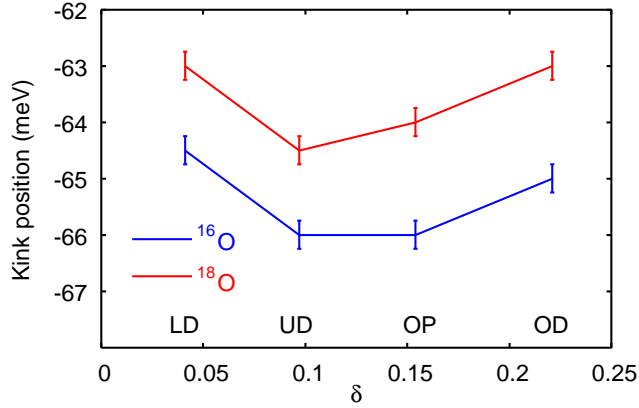}
\caption{\label{figure:3}Doping dependence of kink energies. Bars 
indicate the discretization error during analytic continuation.}
\end{figure}
\begin{figure}
\includegraphics[width=8.4cm]{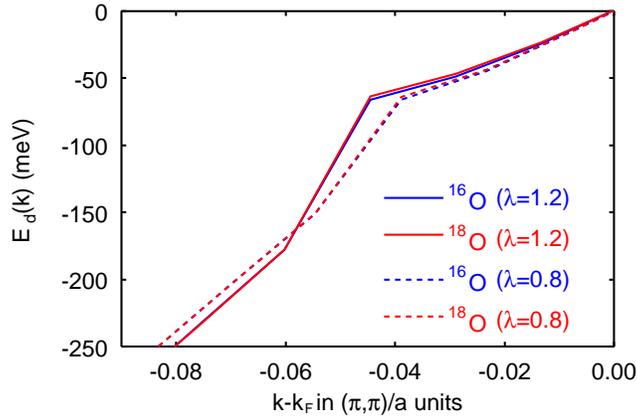}
\caption{\label{figure:4}Dispersion kinks along the nodal direction 
for $\lambda=1.2$. The number of doped holes for 
$^{16}{\mathrm O}(^{18}{\mathrm O})$ material is $\delta=0.1309(0.1310)$. 
The dashed lines show the ones for OP in Table~\ref{table:1}.}
\end{figure}
\begin{figure}
\includegraphics[width=8.4cm]{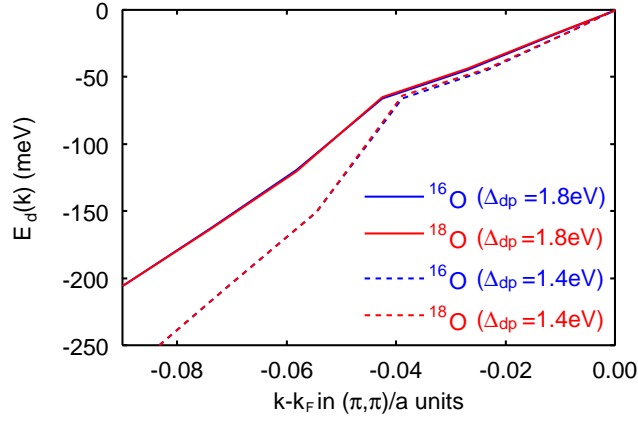}
\caption{\label{figure:5}Dispersion kinks along the nodal direction 
for $\Delta_{dp}=1.8\,{\mathrm{eV}}$. The number of doped holes for 
$^{16}{\mathrm O}(^{18}{\mathrm O})$ material is $\delta=0.1621(0.1622)$. 
The dashed lines show the ones for OP in Table~\ref{table:1}.}
\end{figure}
\begin{figure}
\includegraphics[width=8.4cm]{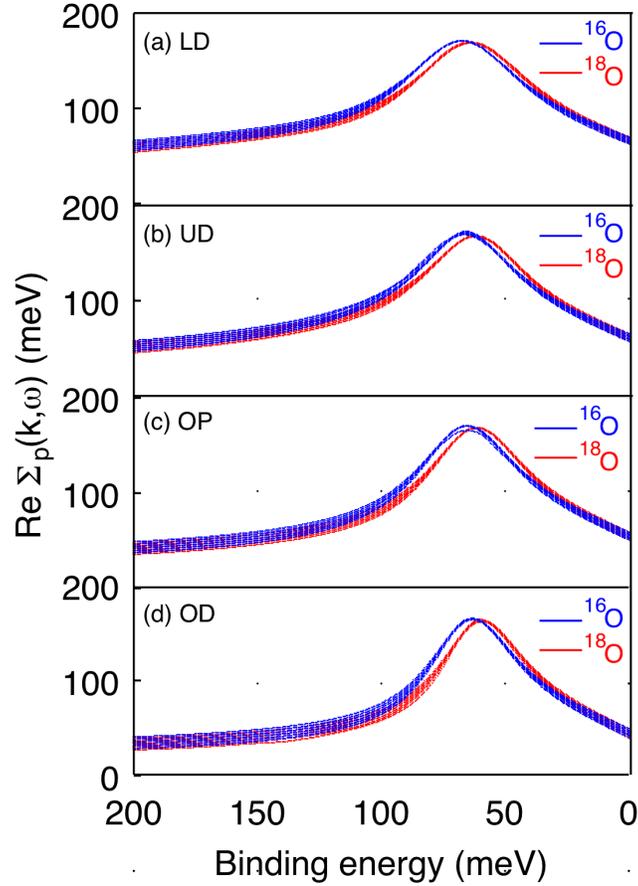}
\caption{\label{figure:6}$\Sigma_p(\mathbf{k},\omega)$ on 
${\mathbf{k}}=\left(\tilde{k}_{\mathrm F}-\frac{n}{64}\right)\!(\pi,\pi)
\hspace{0.5em}(n=0,1,\ldots,5)$, 
where $\tilde{k}_{\mathrm F}=\frac{59}{128}$ for (a), 
$\tilde{k}_{\mathrm F}=\frac{57}{128}$ for (b), 
$\tilde{k}_{\mathrm F}=\frac{55}{128}$ for (c),
and $\tilde{k}_{\mathrm F}=\frac{53}{128}$ for (d), respectively.}
\end{figure}
\begin{figure}
\includegraphics[width=8.4cm]{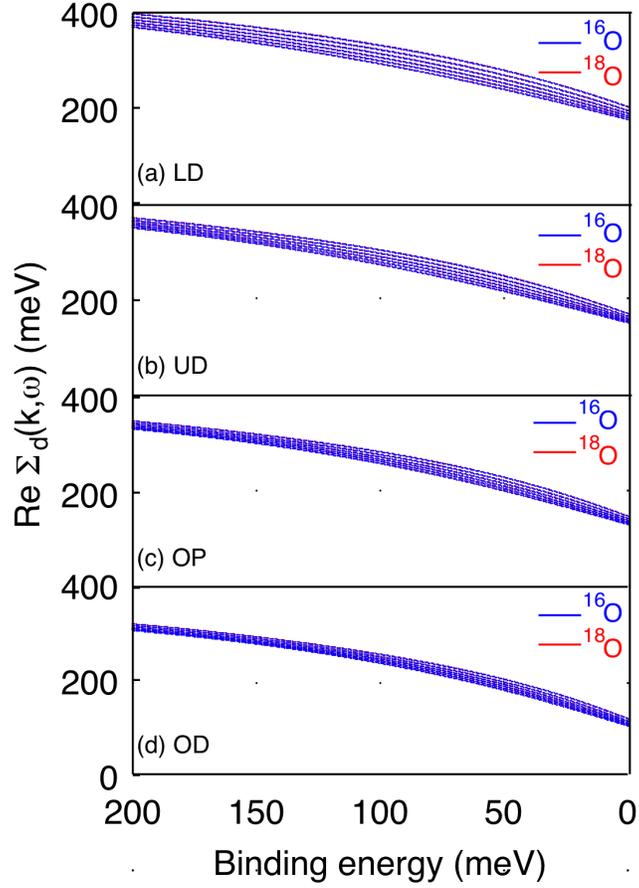}
\caption{\label{figure:7}$\Sigma_d(\mathbf{k},\omega)$ on 
the same ${\mathbf{k}}$ points as in Fig.~\ref{figure:4}.}
\end{figure}
\begin{figure}
\includegraphics[width=8.4cm]{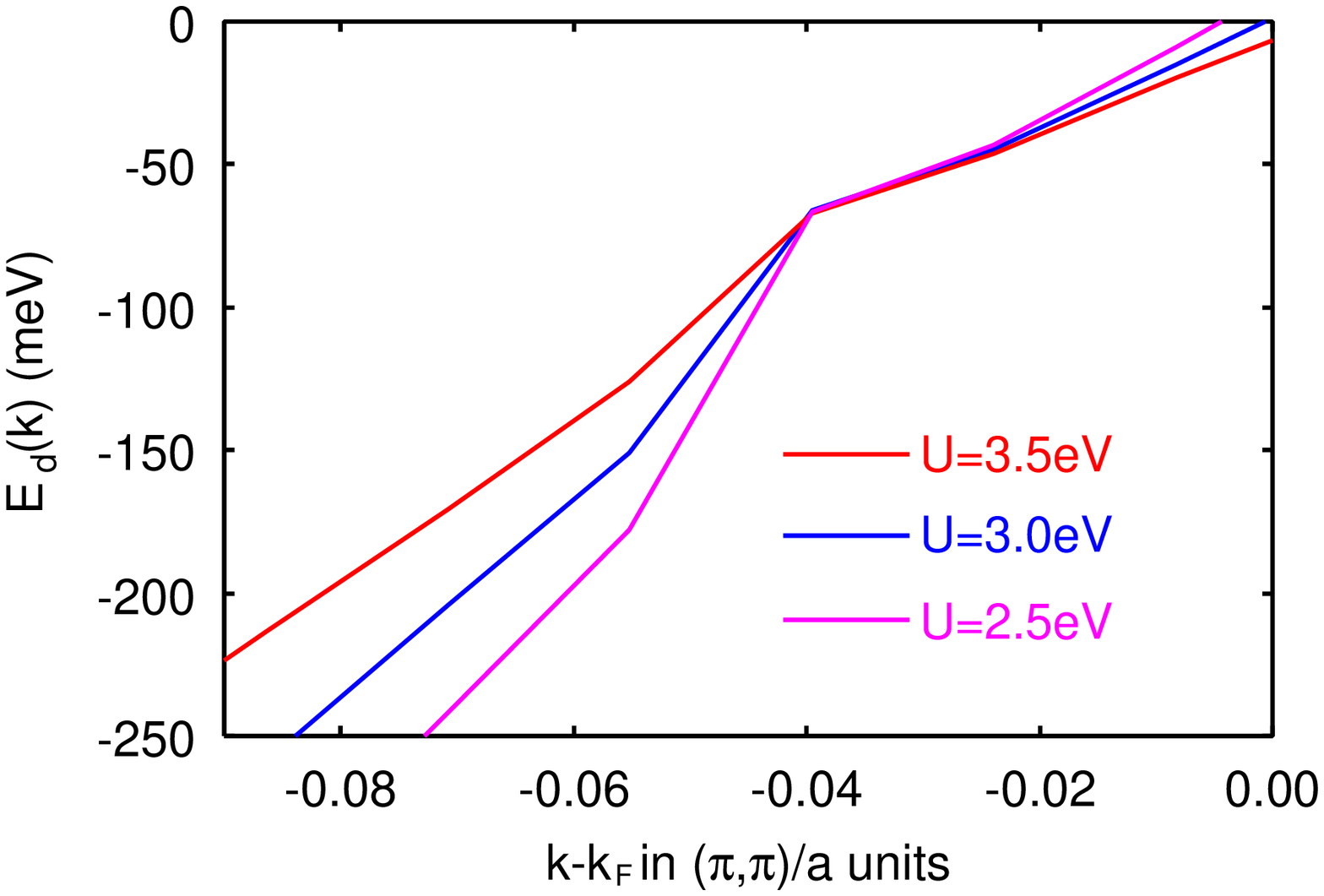}
\caption{\label{figure:8}Dispersion kinks along the nodal direction 
for $U=2.5, 3.0,$ and $3.5\,{\mathrm{eV}}$. They all correspond to 
OP. Momentum is measured from the Fermi surface for 
$U=3.0\,{\mathrm{eV}}$.}
\end{figure}

We show the numbers of doped holes for 
our fully self-consistent solutions in Table~\ref{table:1}. 
The numbers of doped holes 
both for $^{16}{\mathrm O}$ and $^{18}{\mathrm O}$ materials 
are exactly the same to three places of decimals and they 
correspond to four different hole-doped samples, lightly doped (LD),  
underdoped (UD), optimally doped (OP), and overdoped (OD), respectively. 
In Fig.~\ref{figure:1}, we show the color map of the one-particle spectrum 
at Fermi level $A({\mathbf{k}},0)$, where
\begin{equation}
A({\mathbf{k}},\varepsilon) \equiv -\frac{1}{\pi}{\mathrm{Im}}
\left\{{\mathrm{Tr\,}} \mathbf{G}(\mathbf{k},\mathrm{i}\epsilon_n)
\right\}_{\,\mathrm{i}\epsilon_n \rightarrow \varepsilon},
\label{eq:4}
\end{equation}
in order to indicate the Fermi surfaces for $^{16}{\mathrm O}$ materials. 
In Eq.~(\ref{eq:4}), the Pad$\acute{\mathrm{e}}$ approximation is exploited 
for analytic continuation. 
Furthermore, we calculate the 
electronic dispersions of the antibonding band $E_d({\mathbf{k}})$ 
along the nodal direction indicated as the cut in Fig.~\ref{figure:1} 
for all our doping cases. 
$E_d({\mathbf{k}})$ is determined as the $\mathbf{k}$ point on 
which $A({\mathbf{k}},\varepsilon)$ 
has the maximum value at each energy level. Due to this method, 
the curves of $E_d({\mathbf{k}})$ look like a series of line segments, 
as shown in Fig.~\ref{figure:2}. 
We compare every $E_d({\mathbf{k}})$ of our solution with the one 
obtained by another fully self-consistent manner, 
in which the same calculation is performed, except for the EPI. We show 
our results on these dispersions in Fig.~\ref{figure:2}, 
where we can easily recognize that the dispersion kinks along the 
nodal direction appear only when EPI affects the $p^{x(y)}$ electrons. 
The kink energies were slightly shifted by 
$^{16}{\mathrm O}\,\rightarrow\,^{18}{\mathrm O}$ substitution. 
In Fig.~\ref{figure:3}, we detail how these kink energies shift 
depending on hole doping. 
These theoretically evaluated isotope shifts are at most 
$2.5\,{\mathrm {meV}}$, which are much smaller than the ones measured by  
another group's ARPES experiment.~\cite{Gweon04,Gweon06} 
Furthermore, these isotope shifts are almost independent of hole doping 
while another group insists that they are critically affected.~\cite{Gweon07} 
Considering the energy and momentum resolutions in their experiment, 
it may be hard to detect the subtle isotope shifts and their dependence 
on hole doping shown in our model. 

Let us now look at $\lambda$ and $\Delta_{dp}$ dependences in the dispersion 
kinks along the nodal direction in detail. 
Figure~\ref{figure:4} shows the energy 
dispersions for $\lambda=0.8$ (dotted lines) and $\lambda=1.2$ (solid lines). 
There is no clear difference in the isotope effect between 
$\lambda=1.2$ ($2.5\,{\mathrm{meV}}$) and $\lambda=0.8$ 
($2.0\,{\mathrm{meV}}$),
though the dispersion kinks for $\lambda=1.2$ are distinctly shifted 
to the high binding energy compared with those for $\lambda=0.8$. 
On the other hand, 
the isotope shift for $\Delta_{dp}=1.8\,{\mathrm{eV}}$ ($1\,{\mathrm{meV}}$) 
is slightly shrank compared to that for 
$\Delta_{dp}=1.4\,{\mathrm{eV}}$ ($2.5\,{\mathrm{meV}}$), 
though the dispersions depend on the $\Delta_{dp}$ considerably, 
as shown in Fig.~\ref{figure:5}. 
Hence, our model calculation shows that the isotope shifts are not sensitive to 
$\lambda$ and $\Delta_{dp}$. Considering that $\Delta_{dp}$ is closely related 
with EEI in our three band HH model as discussed later, 
we can be fairly certain that the isotope shifts are 
determined by the relative strength between EPI and EEI. 
However, these changes of the isotope shifts are minute, 
thus, our discussions so far are valid regardless 
of $\lambda$ and $\Delta_{dp}$.

To clarify the EPI effect on the $p^{x(y)}$ electrons 
described above, we investigate the $p$ electron self-energy 
$\Sigma_p(\mathbf{k},\omega) \equiv \Sigma_x(\mathbf{k},\omega)+\Sigma_y(\mathbf{k},\omega)$ 
along the nodal direction. In Fig.~\ref{figure:6}, we show 
$\Sigma_p(\mathbf{k},\omega)$ on every six $\mathbf{k}$ points along the 
nodal direction, located inside Fermi surfaces. The energy 
where $\Sigma_p(\mathbf{k},\omega)$ is maximal corresponds to the one 
of the dispersion kink and shifts upward 
by $^{16}{\mathrm O}\,\rightarrow\,^{18}{\mathrm O}$ substitution, 
as shown in Fig.~\ref{figure:2}. The energy dependence of 
$\Sigma_p(\mathbf{k},\omega)$ is definitely due to the EPI introduced with 
the use of Eqs.~(\ref{eq:1}), (\ref{eq:3}), and (\ref{eq:9}). Thus, we can conclude that 
in our solutions for all doping levels from the UD to the OD region, 
the dispersion kinks along the nodal direction are created 
{\textit{only when the EPI is included}}. 

Hereafter, we will discuss why the magnetic ingredients 
hardly bring the dispersion kinks along the nodal direction. 
Even when EPI does not exist, the electrons in our results are exposed 
to the strong AF fluctuation originating from 
the electronic correlation among $d$ electrons. 
In Fig.~\ref{figure:7}, we show $\Sigma_d(\mathbf{k},\omega)$ 
on the same six $\mathbf{k}$ points as in Fig.~\ref{figure:6}. 
It is shown that there is no clear difference in $\Sigma_d(\mathbf{k},\omega)$ 
between $^{16}{\mathrm O}$ and $^{18}{\mathrm O}$ materials. 
The energy dependence of $\Sigma_d(\mathbf{k},\omega)$ 
is definitely due to the electronic correlation introduced with 
the use of Eqs.~(\ref{eq:2})--(\ref{eq:7}). 
We easily recognize that $\Sigma_d(\mathbf{k},\omega)$ uniformly increases 
with the binding energy and has no maximal value up to $200\,{\mathrm{meV}}$ even 
for LD case, in which the strong AF fluctuation is expected to be grown. 
Thus, along the nodal direction, the strong AF fluctuation could cause 
the renormalization of the Fermi velocity, however, it hardly 
promotes any anomalous behavior such as kink structure. 

Finally, we will discuss how our EEI affects the 
dispersion for $^{16}{\mathrm O}$ materials. 
When the on-site Coulomb repulsion $U$ is changed, the Fermi 
velocity is renormalized differently, but this would not change the kink 
energy so much since the kink energy is determined by the EPI alone. 
In Fig.~\ref{figure:8}, we lay out our results for three different $U$s and 
they all correspond to OP. Their total doped holes $\delta$ are slightly 
different as: $\delta = 0.158, 0.154,$ and $0.149$ for $U = 2.5, 3.0,$ 
and $3.5\,{\mathrm{eV}}$, respectively. Hence, 
the Fermi momentum for $U=2.5\,{\mathrm{eV}}$ moves inside 
(or the Fermi surface shrinks) and the one for $U=3.5\,{\mathrm{eV}}$ moves 
outside (or the Fermi surface enlarges), 
compared to the one for $U=3.0\,{\mathrm{eV}}$. These changes of the Fermi 
momenta are small; however, the dispersions at higher energy 
are quite affected, reflecting the binding energy dependence 
of $\Sigma_d(\mathbf{k},\omega)$, as shown in Fig.~\ref{figure:7}. 
Therefore, the dispersion at higher energy could be changed a lot by the EEI
even if the Fermi velocities are almost independent of them. 

\section*{CONCLUSIONS}

By the analysis of our model, we can 
show that the dispersion kink along the nodal direction 
occurs due to EPI. The isotope effect 
upon the electronic dispersion is shown near the 
kink of energy dispersion, not in the high binding energy portions.
~\cite{Gweon04,Gweon06} 
Our evaluation of the subtle isotope shifts has been backed by the report of 
the recent ARPES experiments~\cite{Douglas07,Iwasawa07} 
which show the lack of the unusual isotope effect in the high energy 
portion.~\cite{Gweon04,Gweon06,Gweon07} 
Fortunately for us, our scenario was possibly realized 
in further ARPES experiment.~\cite{Iwasawa08} 
In addition to that, we have investigated how EEI 
effects on the nodal dispersion. It can hardly affect the kink and 
the nodal Fermi velocity; however, it can change the dispersion at higher 
energy. Hence, EPI and EEI play different roles on 
the nodal energy dispersion, respectively. 

Of course, our treatment of EEI is just suited for 
{\textit{weak coupling regime}}, and all of our parameter sets employed 
might be far from the ones for {\textit{strong coupling regime}}. 
If we investigate {\textit{strong coupling regime}} with the use of another 
approach, the low energy structure corresponding to the superexchange $J$ 
could appear in the dispersion. 
However, our results presented here suggest that the structure 
should appear as a broad peak at higher energy due to the frequency dependence 
of the strong AF fluctuation, which will grow into $J$. 
As we all know, other works have already derive qualitatively 
similar conclusions on the basis of other models. 
Some groups adopt $t$-$J$ models
~\cite{Rosch04,Ishihara04,Mishchenko04,Mishchenko06} and other groups do 
one-band HH models.~\cite{Fratini05,Sangiovanni06,Paci06} 
Furthermore, other groups have succeeded in explaining the ARPES 
results.~\cite{Cuk04,Devereaux04,Seibold05,Maksimov05,Meevasana06,Heid07} 
However, in our 2D three-band HH model, both the electron-electron interaction 
among the $d$ electrons and the EPI on $p$ electrons are considered 
according to high-$T_{\mathrm c}$ materials. We believe that it 
is important that quantitatively consistent results with the ARPES 
experiments can be reproduced from such a model. 
The advantage will be when our analysis extends 
to the superconducting state, in which the $p$ electrons play important 
roles as well as $d$ electrons.

\section*{ACKNOWLEDGMENTS}

The authors are grateful to H. Iwasawa and T. Yanagisawa 
for their stimulating discussions. 
The computation in this work was performed on Intel Xeon 
servers at NeRI in AIST.

\end{document}